\documentclass[12pt,letterpaper]{article}

\usepackage[margin=1in]{geometry}
\usepackage{amsmath,amssymb}
\usepackage{physics}
\usepackage{bm}
\usepackage{authblk}
\usepackage{xcolor}
\usepackage{hyperref}
\usepackage{cleveref}
\usepackage{microtype}
\usepackage{setspace}
\usepackage{titlesec}
\usepackage{enumitem}
\usepackage{amsthm}

\usepackage[numbers,sort&compress]{natbib}

\titleformat{\section}{\large\bfseries}{{\Roman{section}.}}{0.5em}{}
\titleformat{\subsection}{\normalsize\bfseries}{{\Alph{subsection}.}}{0.5em}{}

\definecolor{addblue}{rgb}{0.00,0.22,0.72}
\definecolor{delred}{rgb}{0.72,0.09,0.09}
\definecolor{notegray}{rgb}{0.42,0.42,0.42}

\makeatletter
\newsavebox{\qs@dwbox}
\newcommand{\dw}[1]{%
  \sbox{\qs@dwbox}{#1}%
  \leavevmode\rlap{\usebox{\qs@dwbox}}\rule[0.48ex]{\wd\qs@dwbox}{0.5pt}}
\def\qs@delloop#1 #2\qs@delend{%
  \dw{#1}\ifx\relax#2\relax\else\ \qs@delloop#2\qs@delend\fi}
\makeatother

\makeatletter
\@ifundefined{markedversion}{%
  \newcommand{\add}[1]{#1}%
  \newcommand{\del}[1]{}%
  \newcommand{\note}[1]{}%
  \newenvironment{addblock}{}{}%
  \newcommand{\addblk}[1]{#1}%
  \newcommand{\markedlegend}{}%
}{%
  \newcommand{\add}[1]{\textcolor{addblue}{#1}}%
  \newcommand{\del}[1]{\textcolor{delred}{\qs@delloop#1 \relax\qs@delend}}%
  \newcommand{\note}[1]{{\color{notegray}\small\itshape[#1]}}%
  \newenvironment{addblock}{\color{addblue}}{}%
  \newcommand{\addblk}[1]{{\color{addblue}#1}}%
  \newcommand{\markedlegend}{%
    \begin{center}
    \fcolorbox{black!35}{black!4}{%
    \begin{minipage}{0.93\linewidth}
    \small\setstretch{1.0}
    \textbf{Marked copy.}\;
    \textcolor{addblue}{Text in blue is new or rewritten in this revision.}\;
    \textcolor{delred}{\dw{Text} \dw{in} \dw{red} \dw{strikeout} \dw{has} \dw{been} \dw{removed.}}\;
    Changes are shown relative to the manuscript as originally submitted.
    A clean copy is supplied separately.
    \end{minipage}}
    \end{center}
    \vspace{0.5em}}%
}
\makeatother

\newcommand{\bhat}{\hat{\bm{b}}}
\newcommand{\bxi}{\bm{\xi}}
\newcommand{\bE}{\mathbf{E}}
\newcommand{\bB}{\mathbf{B}}
\newcommand{\bA}{\mathbf{A}}
\newcommand{\bJ}{\mathbf{J}}

\newcommand{\bPi}{\bm{\Pi}}
\newcommand{\bv}{\mathbf{v}}
\newcommand{\bx}{\mathbf{x}}
\newcommand{\bX}{\mathbf{X}}
\newcommand{\Astar}{\mathbf{A}^{*}}
\newcommand{\Bstar}{\mathbf{B}^{*}}
\newcommand{\Bparstar}{B^{*}_{\parallel}}
\newcommand{\Hgc}{H_{\mathrm{gc}}}
\newcommand{\Hgcs}{H_{\mathrm{gc},s}}
\newcommand{\Pxi}{P_\xi}
\newcommand{\Pxis}{P_{\xi, s}}
\newcommand{\Lgc}{L_{\mathrm{gc}}}
\newcommand{\fT}{f_{T}}
\newcommand{\ppar}{p_{\parallel}}
\newcommand{\pperp}{p_{\perp}}
\newcommand{\vpar}{v_{\parallel}}

\newcommand{\Omegasc}{\Omega_{s}}
\newcommand{\calH}{\mathcal{H}}
\newcommand{\calF}{\mathcal{F}}
\newcommand{\calA}{\mathcal{A}}
\newcommand{\calQ}{\mathcal{Q}}

\newcommand{\Jact}{\mathcal{J}}
\newcommand{\bkappa}{\bm{\kappa}}
\newcommand{\Jpar}{J_{\parallel}}
\newcommand{\Dani}{\Delta}

\begin{document}

\title{\textbf{A Variational Framework for Guiding-Center Kinetics,
Anisotropic Equilibria, and Quasisymmetry in Stellarators}}

\author[1,*]{Lanke Fu}
\author[2]{Amitava Bhattacharjee}
\affil[1]{Courant Institute School of Mathematics, Computing, and Data Science, New York University, New York, NY}
\affil[2]{Department of Astrophysical Sciences, Princeton University, Princeton, NJ}
\affil[*]{Corresponding author email: lanke\_fu@outlook.com}

\date{}

\maketitle

\markedlegend

\begin{abstract}
We present a variational framework in which (i) guiding-center kinetic theory, (ii) macroscopic force balance with gyrotropic/anisotropic pressure, and (iii) quasisymmetry (QS) constraints appear as different facets of a single structure. Starting from a guiding-center Vlasov--Maxwell action, we \del{performed}\add{perform} constrained variations that \del{yields}\add{yield} the guiding-center kinetic equation and Maxwell equations. Without phenomenological closure, momentum conservation yields macroscopic force balance $\bJ \times \bB/c = \nabla \cdot \bPi$, where $\bJ$ is the current density, $\bB$ is the magnetic field, and $\bPi$ is the gyrotropic stress tensor.
\addblk{Within this framework we recover the classical result that, for any equilibrium distribution built on the invariants of the reduced motion, the parallel projection of this balance
holds identically, so that pressure anisotropy is a kinematic consequence of magnetic-moment conservation together with the variation of $B$ along field lines, rather than a modeling choice.} We connect QS to an integrability condition expressed in coordinate-free form via
\begin{equation*}
  \fT \equiv \nabla\psi \cdot \bigl[\nabla B \times \nabla(\bB \cdot \nabla B)\bigr] = 0,
\end{equation*}
where $\psi$ is the flux surface label\add{, a condition that does not depend on the pressure model}\del{, and show how this condition leads to solvability constraints on anisotropy closely related to those found in a recent constrained Kruskal--Kulsrud variational formulation}.
\add{We then derive, for an arbitrary kinetic equilibrium, an explicit expression for the current crossing the flux surfaces. Its flux-surface average vanishes identically in a quasisymmetric field, but it vanishes pointwise --- as required for Boozer coordinates, and hence for the standard representation of QS --- only if $\ppar+\pperp$ is a flux function. Combined with the parallel identity this forces $\ppar = p(\psi) + K(\psi)B^2$ and $\pperp = p(\psi) - K(\psi)B^2$, consistent with the anisotropic equilibria obtained by E. Rodriguez and A. Bhattacharjee [Phys. Rev. E \textbf{104}, 015213 (2021)] from a constrained Kruskal--Kulsrud functional, in which $(\ppar+\pperp)/2$ is a flux function. Here the structure arises from the kinetic moments without a Lagrange multiplier. For such equilibria the parallel current in a quasisymmetric field follows in closed form, with a single resonance at $\iota = N/M$ in place of the dense set of resonant obstructions of a generic three-dimensional field. Finally, because QS is a Noether symmetry, its breaking is measured by the non-conservation of the associated momentum. We construct a distribution-weighted residual on this basis and relate it to existing optimization metrics.}
\end{abstract}

\section{Introduction}
\label{sec:intro}

Magnetic confinement of high-temperature plasmas relies fundamentally on the structure of particle trajectories in strong magnetic fields. In toroidal geometry, the guiding-center motion of charged particles is governed by the geometry of magnetic field lines and by the symmetries of the magnetic field strength. The existence or absence of such symmetries determines the quality of particle confinement, transport properties, and ultimately the viability of a magnetic confinement configuration.

Among toroidal confinement concepts, stellarators occupy a distinctive position because their confining magnetic fields are intrinsically three-dimensional (3D). Unlike tokamaks, which rely on axisymmetry to guarantee conservation of canonical toroidal momentum, stellarators generally lack continuous geometric symmetry. Guiding-center particle motion in generic stellarator fields is therefore nonintegrable, leading to enhanced neoclassical transport and degraded confinement. A major advance in stellarator theory has been the recognition that a sufficient condition for guiding-center integrability is for the magnetic field strength to possess a hidden symmetry in appropriately chosen magnetic coordinates. This property is known as quasisymmetry (QS) \cite{nuhren1988,boozer1983,rodriguez_necessary_2020}. When such a symmetry exists, guiding-center motion admits an additional conserved quantity analogous to canonical momentum in axisymmetric systems, substantially reducing neoclassical transport\add{. (A quasisymmetric field is in fact isomorphic to an axisymmetric one for the purposes of neoclassical transport \cite{boozer1983}}.) QS has accordingly emerged as a central organizing principle in modern stellarator design and optimization\add{, and underlies configurations such as HSX and NCSX \cite{anderson1995,zarnstorff2001} as well as recent constructions with precise QS \cite{Landreman2022-lm}. The reduction of neoclassical transport by magnetic-field optimization has been demonstrated experimentally in Wendelstein 7-X \cite{beidler2021}}.

In Boozer coordinates $(\psi, \theta, \phi)$, QS requires that the field strength depend on a single helical angle,
\begin{equation}
  B = B(\psi,\, M\theta - N\phi),
  \label{eq:QS_def}
\end{equation}
where $\psi$ labels magnetic flux surfaces and $(M,N)$ are integers. This form ensures the existence of a conserved guiding-center momentum associated with the ignorable coordinate. Despite the success of this geometric formulation, the physical status of QS in relation to the kinetic structure of magnetized plasmas remains incompletely understood. Many treatments \add{consider}\del{introduce} QS as a property of the magnetic field independent of the plasma equation of state \add{\cite{Landreman2022-lm, rodriguez_magnetic, burby2020}}. \add{The associated equilibrium theory} is typically formulated with scalar-pressure \add{\cite{garren_quasihelical_1991, garren_field_strength_1991, Helander2014-ev, qs_opt_example_Paul2021-cy, landreman_nae}, which, historically, is the default in tokamak and stellarator magnetohydrodynamics (MHD)}. However, collisionless magnetized plasmas are intrinsically anisotropic: rapid gyromotion separates parallel and perpendicular dynamics and produces a gyrotropic stress tensor with distinct parallel and perpendicular pressures \cite{cgl1956,kruskal_kulsrud_1958}.

This raises a fundamental question: is the scalar-pressure assumption physically consistent with a quasisymmetric magnetic field? In 3D geometry, the scalar-pressure problem is known to be overdetermined. This is the well-known Garren--Boozer (GB) conundrum \cite{garren_quasihelical_1991, garren_field_strength_1991, landreman_nae, landreman_sengupta_plunk_2019}. We suggest that QS, anisotropic equilibrium, and kinetic invariants are not independent features but facets of a single variational structure. In this paper, we show that if QS is identified rigorously as a Noether symmetry of the guiding-center action, the geometric constraints traditionally imposed by hand emerge naturally as solvability conditions. Crucially, this framework suggests that anisotropic pressure is a structural necessity: it provides the mathematical flexibility required to resolve the question of the existence of 3D equilibria. \del{Because integrability is restored through symmetry, quasisymmetric stellarators are characterized by smooth parallel currents that bypass the singularities in scalar-pressure models.} \add{Because integrability is restored through symmetry, the magnetic differential equation that determines the parallel current in a quasisymmetric field is subject to a single non-resonance condition rather than to the dense set of resonant obstructions encountered in generic 3D geometry \cite{grad1967,newcomb1959,Helander2014-ev}.}

A step toward addressing this question was made recently by Rodriguez and Bhattacharjee, who imposed a coordinate-independent QS condition on the classical Kruskal--Kulsrud variational principle via a Lagrange multiplier, obtaining a class of anisotropic equilibria with a gyrotropic stress tensor \cite{rodriguez_energy_principle}. They also showed via near-axis expansion that a QS equilibrium problem with anisotropic pressure is no longer overdetermined \cite{rodriguez_mhd_1, rodriguez_mhd_2}. These results revealed a striking connection between QS and pressure anisotropy, and suggested that anisotropic equilibrium theory may be more fundamental than scalar-pressure MHD for describing quasisymmetric systems.
From a kinetic perspective, the Vlasov--Maxwell system possesses a rich variational and Hamiltonian structure, with an infinite family of invariants arising from phase-space relabeling symmetry \add{\cite{low1958,morrison1980,marsden_weinstein_1982,morrison1998}}. In strongly magnetized plasmas, guiding-center reduction makes the magnetic moment an adiabatic invariant and reveals the intrinsic gyrotropy of the stress tensor \add{\cite{littlejohn1981,Littlejohn1983-wh,cary_brizard_2009}}. The present work synthesizes these developments within a single unified variational framework.

\begin{addblock}
\add{This paper differs from existing works on quasisymmetric MHD in two key aspects. First, the equilibrium force balance, the pressure tensor, the kinetic equation and the symmetry condition are generated by a single functional. In existing works, QS conditions such as $\fT = 0$ (Eq.~\eqref{eq:fT}) are often treated as imposed constraints on an equilibrium. As discussed above, one example is Ref.~\cite{rodriguez_energy_principle}, where the QS condition is imposed with an external Lagrange multiplier. In this paper, both the QS condition and the force balance follow from guiding-center kinetics without the need for external Lagrange multipliers.}

\add{
Second, Secs.~\ref{sec:parallel_identity} and \ref{sec:parallel_current} connect the kinetic equilibrium to the QS problem explicitly. Section~\ref{sec:parallel_identity} recovers, within the present framework, the classical result of guiding-center equilibrium theory \cite{taylor1963,hall_mcnamara_1975} that the parallel projection of the anisotropic force balance is an identity for any distribution built on the invariants of the reduced motion; we make the treatment of the guiding-center phase-space Jacobian exact. Section~\ref{sec:parallel_current} then derives an explicit expression for the current crossing the flux surfaces in an arbitrary kinetic equilibrium, and shows that the existence of Boozer coordinates restricts the pressures to $\ppar = p(\psi)+K(\psi)B^2$, $\pperp = p(\psi)-K(\psi)B^2$. This is consistent with the constrained Kruskal--Kulsrud formulation of Ref.~\cite{rodriguez_energy_principle}, in which the mean of the parallel and perpendicular pressures is likewise a flux function; here the structure follows from the kinetic moments without a Lagrange multiplier. For this class we obtain the parallel current in closed form.}
\end{addblock}

The remainder of the paper is organized as follows. Section~\ref{sec:variational} reviews the variational structure of collisionless plasma dynamics and the role of dynamically accessible states. Section~\ref{sec:guiding_center} introduces guiding-center reduction and the adiabatic invariants of motion. Section~\ref{sec:action} constructs the unified guiding-center Vlasov--Maxwell action, derives the kinetic and field equations, \add{obtains} macroscopic force balance from Noether's theorem\add{, and shows that the parallel projection of that balance is an identity of the kinetic equilibrium}. Section~\ref{sec:QS} identifies QS as a Noether symmetry of the guiding-center action\add{,} derives the associated geometric integrability constraints\add{, derives the current crossing the flux surfaces in a kinetic equilibrium and the constraint that Boozer coordinates place} on anisotropic pressure\add{, and obtains the parallel current in closed form}. Section~\ref{sec:stability} analyzes the kinetic stability of quasisymmetric equilibria using the energy--Casimir method. Section~\ref{sec:design} discusses implications for equilibrium construction and stellarator optimization\add{, introduces a variational measure of quasisymmetry breaking, and discusses the effect of finite collisionality}. Section~\ref{sec:conclusions} summarizes the principal results and their broader implications.

\section{Variational Principles and Dynamically Accessible States}
\label{sec:variational}

\subsection{Hamiltonian structure of the Vlasov--Maxwell system}

The dynamics of a collisionless plasma is governed by the Vlasov--Maxwell
system, describing the evolution of the distribution function
$f_{s}(\bx,\bv,t)$ for each particle species $s$ together with the
electromagnetic fields. The Vlasov equation is:
\begin{equation}
  \frac{\partial f_{s}}{\partial t}
  + \bv \cdot \nabla f_{s}
  + \frac{q_{s}}{m_{s}}
    \!\left(\bE + \frac{\bv \times \bB}{c}\right)\!
    \cdot \frac{\partial f_{s}}{\partial \bv}
  = 0.
  \label{eq:Vlasov}
\end{equation}
The characteristics of Eq.~\eqref{eq:Vlasov} are \del{HamiltoN's}\add{Hamilton's} equations for the
single-particle Hamiltonian $H_{s} = \tfrac{1}{2}m_{s}v^2 + q_{s}\Phi(\bx)$\add{, with respect to the noncanonical Poisson bracket in $(\bx,\bv)$ that contains the magnetic field (equivalently, of the canonical Hamiltonian $(\mathbf{p}-q_s\bA/c)^2/2m_s + q_s\Phi$)},
so the distribution function is advected by an incompressible phase-space flow
(Liouville's theorem), and the Vlasov equation takes the Poisson-bracket form,
\begin{equation}
  \frac{\partial f_{s}}{\partial t} + \{f_{s}, H_{s}\} = 0,
  \label{eq:Vlasov_PB}
\end{equation}
where the symbols have their usual meanings. The Hamiltonian structure extends to the full Vlasov--Maxwell system with the
Hamiltonian functional
\begin{equation}
\mathcal{H} = \sum_s\int d^3x\,d^3v\; f_s H_s
    + \int d^3x\,\frac{E^2+B^2}{8\pi}.
  \label{eq:Hamiltonian}
\end{equation}

\subsection{Casimir invariants}

The Vlasov--Maxwell system possesses a noncanonical Poisson bracket structure on
functionals of $f_{s}$ and the fields \add{\cite{morrison1980,marsden_weinstein_1982,morrison_greene_1980}}. A key feature is the existence of Casimir
invariants---quantities \add{that Poisson-commute with every functional, and are therefore} conserved independently of the Hamiltonian---arising from
phase-space relabeling symmetry:
\begin{equation}
  C[f_{s}] = \int d^3x\,d^3v\; C(f_{s}),
  \label{eq:Casimir}
\end{equation}
where $C$ is an arbitrary function. These invariants express the preservation of
phase-space topology by Vlasov dynamics.

\begin{addblock}
\add{It is worth recalling where these invariants come from. The Vlasov--Maxwell bracket is degenerate: the Poisson operator possesses a nontrivial kernel, and the functionals annihilated by it are the Casimirs \cite{morrison1980,marsden_weinstein_1982,morrison1998}. The degeneracy is not accidental but is the Eulerian counterpart of a Lagrangian symmetry --- the freedom to relabel the particles that carry the phase fluid, exactly as the analogous relabeling symmetry generates the cross-helicity and magnetic-helicity invariants of ideal magnetohydrodynamics \cite{newcomb1962,morrison_greene_1980}. Because the dynamics is confined to a level set of every Casimir, equilibria are not stationary points of $\calH$ alone but of $\calH$ constrained to such a level set, which is the content of the energy--Casimir principle used below \cite{holm1985,morrison1998,ye_morrison_1992,morrison_pfirsch_1989}.}
\end{addblock}

\subsection{Energy--Casimir variational principle}

Because Vlasov dynamics preserves phase-space topology and measure, admissible
variations of the distribution function must correspond to infinitesimal
canonical transformations,
\begin{equation}
  \delta f_{s} = -\{f_{s}, g_{s}\},
  \label{eq:dyn_acc}
\end{equation}
where $g_{s}(\bx,\bv,t)$ is an arbitrary generating function. These \emph{dynamically accessible} variations preserve all Casimir invariants and represent the kinetic analog of Newcomb's constraint in ideal MHD \add{\cite{newcomb1962,morrison_pfirsch_1990}}. \add{The preservation of Casimir invariants will be used later in Sec.~\ref{sec:stability:casimir} to derive a stability criterion for the equilibrium distribution $f_{0s}$ using the method presented by Morrison \cite{morrison1987variational}.}

Equilibrium states are stationary points of the free-energy functional $\calF = \calH + \sum_{s} C_{s}$. Setting the first variation to zero under constraint \eqref{eq:dyn_acc} yields the condition that the equilibrium distribution \add{$f_{0s}$} depends solely on the constants of motion:
\begin{equation}
  \delta \calF = 0
  \quad \Leftrightarrow \quad
  f_{0s} = F_{s}(I_{1}, I_{2}, \ldots).
  \label{eq:equil_dist}
\end{equation}
\add{Throughout, $f_{0s}$ denotes the time-independent equilibrium guiding-center distribution of species $s$, that is, the solution of $\{f_{0s},\Hgcs\}_{\mathrm{gc}} = 0$, and $F_s$ denotes its functional form when expressed in terms of the invariants.}
\add{Two routes to Eq.~\eqref{eq:equil_dist} should be distinguished. Under the dynamically accessible variations~\eqref{eq:dyn_acc} the Casimirs are held fixed, so $\delta\calF = \delta\calH = \sum_s\int g_s\{f_s,H_s\}$, and stationarity for all $g_s$ gives $\{f_{0s},H_s\}=0$, i.e., $f_{0s}$ is an arbitrary function of the invariants. In the energy--Casimir method, by contrast, the variations of $\calF$ are unconstrained, and the equilibria reached are only those of the form $C'(f_{0s}) = -H_s$, i.e., functions of the energy and of whatever additional quantities the Casimirs may depend on. Invariants that are not Casimirs --- the momentum associated with a symmetry is the relevant example below --- enter only if the corresponding conserved functional is added to $\calF$ \cite{holm1985,morrison1998}. This distinction matters in Sec.~\ref{sec:stability}.}

\section{Guiding-Center Reduction and Invariants of Motion}
\label{sec:guiding_center}

While the Vlasov--Maxwell system provides a complete description of collisionless plasma dynamics, the rapid gyromotion makes it impractical for modeling strongly magnetized plasmas, where the cyclotron frequency $\Omegasc = q_{s}B/(m_{s}c)$ greatly exceeds drift frequencies. \del{The relevant ordering is} \add{Throughout this paper we \emph{assume} the drift-kinetic, or guiding-center, ordering}
\begin{equation}
  \epsilon \equiv \rho/L \ll 1, \qquad k_\perp \rho \sim \epsilon, \qquad \omega/\Omegasc \sim \epsilon ,
  \label{eq:ordering}
\end{equation}
\add{where} $\rho$ is the Larmor radius and $L$ is a macroscopic scale length. \add{This is to be distinguished from the gyrokinetic ordering, in which $k_\perp\rho \sim 1$ is retained while $\omega/\Omegasc \sim \epsilon$ \cite{brizard_hahm_2007,sugama2000}. Finite-Larmor-radius corrections and the gyrophase-dependent part of the action, which are the starting point of any gyrokinetic extension, are outside the scope of the present treatment. We further assume that the plasma is collisionless in the sense that
\begin{equation}
  \nu_{ss'}/\omega_b \ll 1, \qquad \nu_* \ll 1,
  \label{eq:coll_ordering}
\end{equation}
with $\nu_{ss'}$ the collision frequency and $\omega_b$ the bounce frequency, so that the magnetic moment is an adiabatic invariant of the reduced dynamics over the time scales of interest. The consequences of relaxing this last assumption are discussed in Sec.~\ref{sec:collisionality}.}

\subsection{Guiding-center transformation and Lagrangian}

The guiding-center transformation replaces particle coordinates $(\bx, \bv)$ with reduced variables $Z = (\bX, \vpar, \mu, \zeta)$, where $\bX$ is the guiding-center position, $\vpar$ is the velocity parallel to $B$, $\mu = mv_{\perp}^2/(2B)$ is the magnetic moment, a first-order adiabatic invariant in $\epsilon$, and $\zeta$ is the gyrophase. To the lowest order in $\epsilon$, the single-particle guiding-center Lagrangian is \cite{Littlejohn1983-wh}\add{, see also Refs.~\cite{littlejohn1981,cary_brizard_2009}}:
\begin{equation}
  \label{eq:Lgc}
 \Lambda_\text{gc} = \frac{q}{c}\,\Astar(\bX, \vpar) \cdot \dot{\bX}+\frac{mc}{q}\mu\dot\zeta
       - \Hgc,
\end{equation}
with
\begin{equation}
  \Astar = \bA + \frac{mc}{q}\vpar\,\bhat,
  \qquad
  \Hgc = \tfrac{1}{2}m\vpar^2 + \mu B + q\Phi.
  \label{eq:Astar_Hgc}
\end{equation}
Here $\bhat = \bB/B$. The equations of motion follow from the guiding-center Poisson bracket $\dot{Z}^{\alpha} = \{Z^{\alpha}, \Hgc\}_{\mathrm{gc}}$, where
\begin{equation}
  \{F, G\}_{\mathrm{gc}}
  = \frac{\Bstar}{mB_{\parallel}^{*}} \cdot
    \!\left(\nabla F\,\frac{\partial G}{\partial \vpar}
          - \nabla G\,\frac{\partial F}{\partial \vpar}\right)
  + \frac{c}{q\Bparstar}\,\bhat \cdot (\nabla F \times \nabla G),
  \label{eq:PB_gc}
\end{equation}
$\Bstar = \bB + (mc/q)\vpar \nabla\times\bhat$, and $\Bparstar = \Bstar \cdot \bhat$.

\subsection{Adiabatic invariants and equilibrium distribution structure}

The guiding-center dynamics conserves the energy $\Hgc$ in static fields and the magnetic moment $\mu$. \del{Conservation of $\mu$ directly implies pressure anisotropy in macroscopic plasma behavior.} \add{Conservation of $\mu$ is the origin of pressure anisotropy in macroscopic plasma behavior. The mechanism is the variation of the field strength along a field line: since $\mu$ is invariant while $\Hgc$ is conserved, a particle moving into a region of larger $B$ converts parallel into perpendicular energy through the mirror force $-\mu\nabla_\parallel B$, and the accessible range of $\mu$ at fixed position is itself a function of $B$ through the trapping condition. The parallel and perpendicular moments of any distribution built on $(\Hgc,\mu)$ therefore differ wherever $B$ varies along $\bB$. We make this statement quantitative in Sec.~\ref{sec:parallel_identity} by showing that $\partial \ppar/\partial B|_\psi = (\ppar-\pperp)/B$ holds identically for such distributions.} Additional invariants arise with continuous symmetry: in axisymmetric systems, invariance under toroidal rotation conserves the canonical toroidal momentum $P_{\phi} = (q/c)A_{\phi} + m\vpar b_{\phi}$.

In general 3D fields, no additional invariant exists beyond energy and magnetic moment, guiding-center motion is nonintegrable, and particle orbits drift radially. QS is the special condition under which a third invariant exists even in a 3D field: the single-helicity dependence of $B$ creates an effective symmetry of $\Hgc$, restoring integrability on magnetic surfaces.

From Eq.~\eqref{eq:equil_dist}, equilibrium distributions depend only on the invariants of motion. Accordingly, the presence or absence of additional invariants determines the functional form of $f_{0s}$ and the structure of macroscopic force balance. Conservation of $\mu$ implies anisotropic pressure, while the additional QS invariant imposes further constraints on the distribution function.

\section{Unified Eulerian Guiding-Center Vlasov--Maxwell Action}
\label{sec:action}

Before constructing the action, we perform the gyrophase average that reduces
the six-dimensional guiding-center phase space $(\bX, \vpar, \mu, \zeta)$ to
the five-dimensional reduced phase space $(\bX, \vpar, \mu)$.  This step is
justified by the observation that $\Hgc$ is independent of the gyrophase $\zeta$ at leading order.
Consequently, $\zeta$ evolves as a fast, decoupled angle and the distribution
function is gyrophase-independent to leading order:
$f_s(\bX,\vpar,\mu,\zeta,t) \approx \bar{f}_s(\bX,\vpar,\mu,t)$, where
$\bar{f}_s = \add{\int_0^{2\pi}} f_s\,d\zeta/(2\pi)$ is the gyrophase average. \add{For brevity, we will drop the overbar in $\bar f_s$ as simply $f_s(\bX,\vpar,\mu,t)$  in the rest of the paper. We will also drop the gyrophase $\zeta$ from the reduced variables $Z = (\bX, \vpar, \mu)$.}
The Berry phase term $(mc/q)\mu\dot{\zeta}$ in the Lagrangian encodes the holonomy of the gyromotion
and is responsible for the noncanonical structure of the guiding-center Poisson
bracket. After gyrophase averaging it drops out, leaving the
reduced Lagrangian \cite{Littlejohn1983-wh}
\begin{equation}
  \bar{\Lambda}_{\mathrm{gc}} = \frac{q}{c}\,\bA^{*}(\bX,\vpar)\cdot\dot{\bX}
    - \Hgc.
\end{equation}
The gyrophase-averaged guiding-center action with invariant phase-space measure
\del{$d^5Z=$ $\Bparstar\,d^3X$ $d\vpar\,d\mu_s$}\add{$d^5Z = (2\pi/m_s)\,\Bparstar\,d^3X\,d\vpar\,d\mu_s$ (the gyrophase integral having been performed, and the factor $2\pi/m_s$ chosen so that $d^5Z$ is the physical phase-space volume $d^3X\,d^3v$ to leading order)} was
formulated by Brizard \& Tronci \cite{Brizard2016-ts}, whose derivation of
the Euler--Lagrange equations and Noether force balance we follow\add{; see also Ref.~\cite{pfirsch_morrison_1985} for local conservation laws in guiding-center theory, and related Hamiltonian and Euler--Poincar\'e formulations are given in Refs.~\cite{squire2013,burby2015,sugama2000}}. All results
in Sections~\ref{sec:action}--\ref{sec:design} operate at this
gyrophase-averaged level; the full $\zeta$-dependent action is the natural
starting point for gyrokinetic extensions involving finite-Larmor-radius
corrections, which are deferred to future work.

\subsection{Action functional}

We consider a plasma with guiding-center distribution functions $f_{s}(Z,t)$
and electromagnetic potentials $(\Phi, \bA)$. The total action is
\begin{equation}
  \calA = \int dt\,\bigl[L_{\mathrm{gc}} + L_{\mathrm{field}}\bigr],
  \label{eq:action}
\end{equation}
with
\begin{gather}
  L_{\mathrm{gc}}
    = \sum_{s} \int d^5Z\,\Lambda_{\text{gc},s}(Z,\dot{Z})\,f_s(Z),
  \qquad
  \label{eq:L_parts1}\\
  L_{\mathrm{field}}
    = \int d^3x\,\frac{E^2 - B^2}{8\pi}.
  \label{eq:L_parts2}
\end{gather}
Here, the phase-space volume element is the invariant guiding-center measure, \del{$\sum_s\int{d^5Z}$ $\equiv$ $\sum_s\int\Bparstar\,d^3X$ $d\vpar\,d\mu_s$}\add{$\sum_s\int d^5Z \equiv \sum_s(2\pi/m_s)\int\Bparstar\,d^3X\,d\vpar\,d\mu_s$}, ensuring the conservation of phase-space volume.

Admissible variations of the distribution function are restricted to the dynamically accessible form~\eqref{eq:dyn_acc}:
\begin{equation}
  \delta f_{s} = -\{f_{s}, g_{s}\}_{\mathrm{gc}},
  \label{eq:dyn_acc_gc}
\end{equation}
where $g_{s}(Z,t)$ is an arbitrary generating function.

\subsection{Euler--Lagrange equations}
Brizard and Tronci \cite{Brizard2016-ts} showed that the stationarity of the action yields the guiding-center Vlasov equation. Vary Eq. \eqref{eq:action} with respect to $f_{s}$ and integrate it by parts in phase space; the Euler--Lagrange equation is:
\begin{equation}
  \frac{\partial f_{s}}{\partial t}
  + \{f_{s}, \Hgcs\}_{\mathrm{gc}} = 0.
  \label{eq:gc_Vlasov}
\end{equation}
Variation with respect to the scalar potential $\Phi$ yields Gauss's law,
$\nabla \cdot \bE = 4\pi\rho_e$, with
\begin{equation*}
  \rho_e(\bx,t)
  = \sum_{s} q_{s} \int d^5Z\, f_s\,\delta(\bx - \bX).
\end{equation*}
Variation with respect to the vector potential $\bA$ yields Ampere's law,
\begin{equation}
  \nabla \times \bB - \frac{1}{c}\frac{\partial \bE}{\partial t}
  = \frac{4\pi}{c}\bJ,
  \label{eq:Ampere}
\end{equation}
with
\begin{equation}
    \bJ(\bx,t) = \sum_{s} q_{s} \int d^5Z\, f_s\dot{\bX}\,\delta(\bx - \bX)
    \addblk{\; + \; c\,\nabla\times\mathbf{M},
    \qquad
    \mathbf{M}(\bx,t) = -\sum_s \int d^5Z\, f_s\,\mu_s\,\bhat\,\delta(\bx-\bX)}.
\end{equation}
\add{The second term is the magnetization current, which arises from the dependence of $\Hgc$ on $B = |\nabla\times\bA|$ through $\mu B$. The dependence of $\Astar$ on $\bhat$, and of the Jacobian $\Bparstar$ on the field, produce further contributions of higher order in $\epsilon$ associated with parallel flows; the complete expressions are given in Ref.~\cite{Brizard2016-ts}. It is the total current, guiding-center plus magnetization, that enters the force balance below.}
The coupled guiding-center Vlasov--Maxwell system thus follows entirely from
the unified action as the Euler--Lagrange equations.

\subsection{Macroscopic force balance from Noether's theorem}

By Noether's theorem, the invariance of the action under spatial translations
$\bx \to \bx + \epsilon\bxi$ ($\bxi$ constant) yields the conservation of total
momentum. For brevity, define the ``velocity''-space measure \del{$d^2V\equiv\Bparstar{d\vpar}d\mu_s$}\add{$d^2V\equiv (2\pi/m_s)\,\Bparstar\, d\vpar\, d\mu_s$}, so that $d^5Z = d^2Vd^3X$\add{, $\int d^2V f_s = n_s$ is the density of species $s$, and the $\vpar$ integral runs over both signs of $\vpar$. This normalization is used throughout}. The total momentum density is $\mathbf{P} = \mathbf{P}_{p} +
\mathbf{P}_{f}$, with the particle and field contributions
\begin{equation}
  \mathbf{P}_{p}
    = \sum_{s} \int d^2V\, f_{s}\,m_s\vpar\,\bhat,
  \qquad
  \mathbf{P}_{f}
    = \frac{\bE \times \bB}{4\pi c}.
  \label{eq:momenta}
\end{equation}
Equations (71)-(80) of Brizard and Tronci \cite{Brizard2016-ts} show that applying Noether's theorem gives the momentum conservation law
\begin{equation}
  \frac{\partial \mathbf{P}}{\partial t} + \nabla \cdot \bPi_\text{tot} = 0,
  \label{eq:mom_cons}
\end{equation}
where $\bPi_\text{tot}=\bPi_p + \bPi_f$ is the total stress tensor. The field contribution $\bPi_f$ is
\begin{equation}
    \bPi_f=\frac{\mathbf{I}}{8\pi}\left(E^2 + B^2\right)-\frac{1}{4\pi}\left(\bE\bE+\bB\bB\right),
\end{equation}
and the particle contribution $\bPi_p$ is
\begin{equation}
  \bPi_{p}
  = \sum_{s} \int d^2V\, f_{s}\,\left[m_s
      \vpar^2\,\bhat\bhat
      + \mu_s B\,(I - \bhat\bhat)
    \right]\!.
  \label{eq:stress_tensor}
\end{equation}
Define the gyrotropic pressure tensor with this contribution,
\begin{equation}
  \bPi = \pperp I + (\ppar - \pperp)\,\bhat\bhat,
  \label{eq:gyrotropic_P}
\end{equation}
where the parallel and perpendicular pressures are
\begin{equation}
  \ppar = \sum_{s} m_s   \int d^2V\, f_{s}\,\vpar^2,
  \qquad
  \pperp = \sum_{s}   \int d^2V\, f_{s}\,\mu_s B.
  \label{eq:pressures}
\end{equation}
For stationary equilibria, Eq.~\eqref{eq:mom_cons} reduces to $\nabla \cdot \bPi_\text{tot} = 0$. Separating the particle and field stresses and applying quasi-neutrality gives the force-balance equation $\nabla \cdot \bPi -\bJ \times \bB/c=0$. Substituting in Eq.~\eqref{eq:gyrotropic_P} yields the Chew--Goldberger--Low (CGL) force-balance equation \add{\cite{cgl1956}},
\begin{equation}
  \nabla\pperp
  + \nabla \cdot \left[(\ppar - \pperp)\,\bhat\bhat\right]
 - \frac{\bJ \times \bB}{c}=0
  .
  \label{eq:CGL}
\end{equation}
Note that the pressure anisotropy arises directly from the guiding-center distribution function and requires no phenomenological closure.

\begin{addblock}
\subsection{Parallel force balance is an identity of the kinetic equilibrium}
\label{sec:parallel_identity}

\add{The parallel projection of Eq.~\eqref{eq:CGL} makes explicit the kinetic origin of the anisotropy. Writing $\Dani \equiv \ppar - \pperp$ and using $\nabla\cdot\bhat = -\bhat\cdot\nabla\ln B$, which follows from $\nabla\cdot\bB = 0$, together with $\bhat\cdot\bkappa = 0$ for the curvature $\bkappa = (\bhat\cdot\nabla)\bhat$, the divergence of the gyrotropic stress may be written}
\begin{equation}
  \nabla\cdot\bPi = \nabla \pperp + B\,\bhat\,\bhat\cdot\nabla\!\left(\frac{\Dani}{B}\right) + \Dani\,\bkappa .
  \label{eq:divPi}
\end{equation}
\add{Since $\bhat\cdot(\bJ\times\bB) = 0$, the parallel projection of the force balance gives the familiar CGL parallel balance}
\begin{equation}
  \bB\cdot\nabla \ppar = \left(\ppar - \pperp\right)\bB\cdot\nabla \ln B .
  \label{eq:parbalance}
\end{equation}

\add{Equation~\eqref{eq:parbalance} is an identity satisfied by the moments of any distribution built on the invariants of the reduced motion. This is a classical result of guiding-center equilibrium theory \cite{taylor1963,hall_mcnamara_1975,grad1967}, usually stated as $\pperp = \ppar - B\,\partial\ppar/\partial B$.}

\add{Let $f_{0s} = F_s(\Hgcs,\mu_s)$. With the measure $d^2V = (2\pi/m_s)\Bparstar\,d\vpar\,d\mu_s$ of Sec.~\ref{sec:action}, the moments~\eqref{eq:pressures} of species $s$ are}
\begin{equation}
  p_{\parallel s} = 2\pi \!\int\!\! d\mu_s\!\int_{-\infty}^{\infty}\!\! d\vpar\; \Bparstar F_s\,\vpar^{2},
  \qquad
  p_{\perp s} = \frac{2\pi B}{m_s} \!\int\!\!  d\mu_s \!\int_{-\infty}^{\infty}\!\! d\vpar \; \Bparstar F_{s}\,\mu_s ,
  \label{eq:moments_mu}
\end{equation}
\add{with $\ppar = \sum_s p_{\parallel s}$ and $\pperp = \sum_s p_{\perp s}$. The Jacobian is}
\begin{equation}
  \Bparstar = B + \frac{m_s c}{q_s}\,\vpar\,\bhat\cdot\nabla\times\bhat .
  \label{eq:Bparstar_parity}
\end{equation}
\add{Its second term depends on the field-line torsion $\bhat\cdot\nabla\times\bhat$, which is not a function of $(\psi,B)$, so that if it survived in Eq.~\eqref{eq:moments_mu} the pressures would not be functions of $(\psi,B)$ alone. It does not survive: $F_s(\Hgcs,\mu_s)$ depends on $\vpar$ only through $\vpar^2$, both integrands in Eq.~\eqref{eq:moments_mu} are even in $\vpar$ apart from the second term of Eq.~\eqref{eq:Bparstar_parity}, which is odd, and that term therefore integrates to zero over the two signs of $\vpar$. The replacement $\Bparstar \to B$ is thus exact for these moments. Changing the integration variable from $\vpar > 0$ to $\Hgcs$ at fixed $\mu_s$, with $m_s\vpar\,d\vpar = d\Hgcs$ and}
\begin{equation}
  \vpar = \left[\frac{2}{m_s}\left(\Hgcs - \mu_s B - q_s\Phi\right)\right]^{1/2},
  \label{eq:vpar_of_H}
\end{equation}
\add{and accounting for both signs of $\vpar$ by a factor of 2, the moments become}
\begin{equation}
  p_{\parallel s} = \frac{4\pi B}{m_s}
  \!\int\!\! d\mu_s\!\int_{\mu_s B + q_s\Phi}^{\infty}\!\!\!\! d\Hgcs\;
  F_s\,\vpar ,
  \qquad
  p_{\perp s} = \frac{4\pi B^{2}}{m_s^{2}}
  \!\int\!\! d\mu_s\!\int_{\mu_s B + q_s\Phi}^{\infty}\!\!\!\! d\Hgcs\;
  \frac{F_s\,\mu_s}{\vpar}.
  \label{eq:moments_H}
\end{equation}
\add{Differentiate $p_{\parallel s}$ with respect to $B$ at fixed $\psi$ and $\Phi$. The distribution $F_s$ does not depend on $B$; the contribution from the lower limit of integration vanishes because $\vpar \to 0$ there; the explicit factor $B$ gives $p_{\parallel s}/B$; and from Eq.~\eqref{eq:vpar_of_H}, $\partial \vpar/\partial B|_{\Hgcs,\mu_s} = -\mu_s/(m_s\vpar)$. Hence}
\begin{equation}
  \left.\frac{\partial p_{\parallel s}}{\partial B}\right|_{\psi,\Phi}
  = \frac{p_{\parallel s}}{B} - \frac{4\pi B}{m_s^{2}}\!\int\!\! d\mu_s\!\int\!\! d\Hgcs\;\frac{F_s\,\mu_s}{\vpar}
  = \frac{p_{\parallel s} - p_{\perp s}}{B}.
  \label{eq:CGLidentity}
\end{equation}
\add{The identity holds species by species and therefore for the summed pressures. If $\Phi$ is a flux function, $\ppar$ and $\pperp$ depend on position only through $\psi$ and $B$, so that $\bB\cdot\nabla \ppar = (\partial \ppar/\partial B)\,\bB\cdot\nabla B$, and Eq.~\eqref{eq:parbalance} is satisfied identically.}

\add{The assumption that $\Phi$ is a flux function is not needed, and is in general not consistent with anisotropy, since quasineutrality between species with different anisotropies usually requires a potential that varies along the field. The same calculation gives $\partial p_{\parallel s}/\partial\Phi|_{\psi,B} = -q_s n_s$, so that for each species}
\begin{equation}
  \bB\cdot\nabla p_{\parallel s} = \left(p_{\parallel s} - p_{\perp s}\right)\bB\cdot\nabla \ln B - q_s n_s\,\bB\cdot\nabla\Phi ,
  \label{eq:parbalance_species}
\end{equation}
\add{which is the complete parallel momentum balance of species $s$, electric force included. Summing over species, the electric term is $-\rho_e\,\bB\cdot\nabla\Phi$, which vanishes under quasineutrality, and Eq.~\eqref{eq:parbalance} again follows identically.}

\add{Three consequences follow. First, Eq.~\eqref{eq:CGLidentity} is the precise content of the statement, made qualitatively in Sec.~\ref{sec:guiding_center}, that conservation of $\mu$ implies pressure anisotropy: the anisotropy \emph{is} the local response of the parallel pressure to the variation of the field strength along a field line. Second, pointwise isotropy requires $\partial\ppar/\partial B|_\psi = 0$ over the range of $B$ on each surface. This holds in particular when $F_s$ is independent of $\mu_s$, i.e., a function of $\Hgcs$ alone. Since $\mu_s$ is an independent adiabatic invariant of the reduced dynamics, and since trapping makes the accessible range of $\mu_s$ depend on position along the field line, isotropy is generically a restriction on the distribution function rather than the typical collisionless case. Third, no closure has been invoked anywhere: the parallel force balance is a property of the moments of an arbitrary function of the invariants. This is the precise sense in which the anisotropic force balance is derived rather than assumed.}

\add{When the additional quasisymmetric invariant is retained, $f_{0s} = F_s(\Hgcs,\mu_s,\Pxis)$, the argument goes through at leading order in $\epsilon$. From Eq.~\eqref{eq:Pxi} below, $\Pxi = (q/c)\bA\cdot\bxi + m\vpar(\bhat\cdot\bxi)$, and the first term is a flux function in Boozer coordinates, so that $\Pxis = P_0(\psi) + O(\epsilon)$ and $\ppar,\pperp$ remain functions of $(\psi,B)$ up to $O(\epsilon)$ corrections. The second term of $\Pxi$ is odd in $\vpar$; through it the distribution acquires an odd part, the cancellation of the second term of Eq.~\eqref{eq:Bparstar_parity} is no longer exact, and the resulting $O(\epsilon)$ corrections carry the parallel flows and the departure from Eq.~\eqref{eq:CGLidentity}.}
\end{addblock}

\section{Quasisymmetry as a Noether Symmetry}
\label{sec:QS}

We now show that QS arises naturally as a continuous symmetry of the
guiding-center action. Consider a transformation of guiding-center coordinates
generated by a vector field $\bxi(\bX)$,
\begin{equation}
  \delta \bX = \epsilon\,\bxi(\bX),
  \quad
  \delta\vpar = 0,
  \quad
  \delta\mu = 0,
  \label{eq:sym_transf}
\end{equation}
where $\epsilon$ is infinitesimal. The action is invariant under
\eqref{eq:sym_transf} if the guiding-center Lagrangian changes only by a total
time derivative, $\delta \Lgc = d\mathcal{S}/dt$. Invariance of the guiding-center
Hamiltonian~\eqref{eq:Astar_Hgc} under \eqref{eq:sym_transf} requires
\begin{equation}
  \bxi \cdot \nabla B = 0,
  \qquad
  \bxi \cdot \nabla\Phi = 0.
  \label{eq:QS_cond}
\end{equation}
Of these two conditions, we focus our attention on the first, since the second condition, potential invariance, is generically satisfied on flux surfaces in MHD equilibria. \add{Invariance of $\Hgc$ is not, however, sufficient. Equation~\eqref{eq:QS_cond} together with tangency to the flux surfaces, $\bxi\cdot\nabla\psi = 0$, is satisfied by $\bxi \propto \nabla\psi\times\nabla B$ in \emph{any} field with nested surfaces, and so by itself places no restriction on the field. The content of QS lies in the invariance of the symplectic part $(q/c)\Astar\cdot\dot{\bX}$ of the Lagrangian, which requires $\mathcal{L}_{\bxi}\Astar$ to be exact. At leading order in $\epsilon$ this gives \cite{burby_qin_2013,rodriguez_necessary_2020,burby2020}}
\begin{equation}
  \addblk{\nabla\cdot\bxi = 0, \qquad \bB\times\bxi = \nabla\Phi_\xi ,}
  \label{eq:QS_sympl}
\end{equation}
\add{where $\mathcal{L}_{\bxi}\bA = \bxi\times\bB + \nabla(\bxi\cdot\bA)$ has been used. Equations~\eqref{eq:QS_cond} and \eqref{eq:QS_sympl} together are the weak QS conditions of Ref.~\cite{rodriguez_necessary_2020}, where it is also shown that they imply the existence of flux surfaces, labeled by $\Phi_\xi$ (so that we may take $\Phi_\xi = \psi$ up to a flux-function factor), regardless of whether the field is in any equilibrium. Note that the term $m\vpar\bhat$ in $\Astar$ is not exactly invariant under such a $\bxi$; this is the reason that QS is a property of the leading-order guiding-center dynamics.} In axisymmetric systems, $B$ is invariant under toroidal rotations. In 3D fields, \del{the same condition defines} \add{these conditions define} QS: the field strength
depends on a single helical coordinate. When the action is invariant, Noether's
theorem yields the conserved quantity
\begin{equation}
  \Pxi = \frac{\partial \Lambda_\text{gc}}{\partial \dot{\bX}} \cdot \bxi
           = \frac{q}{c}\,\Astar \cdot \bxi,
  \label{eq:Pxi}
\end{equation}
which generalizes the canonical momentum associated with geometric symmetry and restores the integrability of particle trajectories on magnetic surfaces. From Eq.~\eqref{eq:equil_dist}, equilibrium distributions in quasisymmetric systems take the form
\begin{equation}
  f_{0s} = F_s(\Hgcs, \mu_s, \Pxis).
  \label{eq:equil_QS}
\end{equation}
Because $f_{0s}$ depends on $\Pxis$ in addition to $\Hgcs$ and $\mu_s$, the pressure tensor acquires additional structure determined by magnetic geometry. This directly connects QS to anisotropic pressure and macroscopic equilibrium.

\subsection{Solvability condition and the geometric constraint}
\label{sec:solvability}

Since $\bxi$ must be tangent to magnetic surfaces to preserve flux labeling, we require $\bxi\cdot\nabla\psi = 0$. Together with \eqref{eq:QS_cond}, we thus have three conditions on the vector field $\bxi$. These conditions imply that $\bxi$ lies within the flux surface and is perpendicular to $\nabla B$ within that surface.  Geometrically, $\bxi$ must therefore point along the \emph{level curves of $B$ on the flux surface}, i.e., $\bxi \propto \nabla\psi\times\nabla B$ (assuming $\nabla\psi$ and $\nabla B$ are linearly independent).  In Boozer coordinates $(\psi,\theta,\phi)$, this means $B = B(\psi, M\theta - N\phi) \equiv B(\psi,\chi)$, where $\chi$ is the conjugate to the ignorable coordinate $\bxi = N\partial_\theta + M\partial_\phi$ (up to normalization). \add{Two qualifications are needed. First, tangency to the flux surfaces is not an independent requirement: it follows from Eq.~\eqref{eq:QS_sympl}, since $\bxi\cdot\nabla\Phi_\xi = \bxi\cdot(\bB\times\bxi) = 0$. Second, it is Eq.~\eqref{eq:QS_sympl}, not Eq.~\eqref{eq:QS_cond}, that singles out $\bxi = N\partial_\theta + M\partial_\phi$ among the vector fields tangent to the level curves of $B$; and the use of Boozer coordinates presupposes that they exist, which requires $\bJ\cdot\nabla\psi = 0$. In the anisotropic case this is a nontrivial condition on the equilibrium, examined in Sec.~\ref{sec:parallel_current}.}

The conditions on $B$ established so far --- $B = B(\psi,\chi)$ --- are sufficient for the Noether invariant $\Pxi = (q/c)\bA^*\cdot\bxi$ to exist as a conserved guiding-center momentum. \del{They are \emph{not} sufficient, however, to guarantee} \add{In Boozer coordinates they also guarantee} that the second adiabatic invariant $\Jact=\oint m v_\parallel d\ell$ is a flux function, i.e., that bounce-averaged drift orbits remain on flux surfaces\add{, because $\bB\cdot\nabla B = B^2(\iota M-N)\,\partial_\chi B/(G+\iota I)$ is then automatically a function of $(\psi,\chi)$; QS implies omnigenity \cite{cary_shasharina_1997,landreman_catto_2012}. When Boozer coordinates are not assumed, the corresponding requirement must be stated separately}.  \del{For $\Jact$ to be well-defined as a global adiabatic invariant, one needs the additional condition} \add{It is the condition} that $\bB\cdot\nabla B$ is also ignorable under $\bxi$:
\begin{equation}
  \bxi\cdot\nabla(\bB\cdot\nabla B) = 0,
  \label{eq:BgradB_sym}
\end{equation}
i.e., $\bB\cdot\nabla B = \mathcal{G}(\psi,\chi)$ for some function $\mathcal{G}$. The coordinate-free expression of the combined conditions --- $B = B(\psi,\chi)$ together with \eqref{eq:BgradB_sym} --- is the triple-product condition:
\begin{equation}
  \fT \equiv \nabla\psi\cdot\bigl[\nabla B\times\nabla(\bB\cdot\nabla B)\bigr] = 0.
  \label{eq:fT}
\end{equation}
To see why: when $B = B(\psi,\chi)$ and $\bB\cdot\nabla B = \mathcal{G}(\psi,\chi)$, both $\nabla B$ and $\nabla(\bB\cdot\nabla B)$ lie in the two-dimensional span of $\{\nabla\psi, \nabla\chi\}$, so their cross product is proportional to $\nabla\psi
\times\nabla\chi$ and the triple-product $\fT = \nabla\psi\cdot(\nabla B\times
\nabla(\bB\cdot\nabla B))$ vanishes identically. Conversely, $\fT = 0$ implies these two conditions \cite{rodriguez_necessary_2020}\add{ (see also Ref.~\cite{rodriguez_magnetic})}.

\begin{addblock}
\add{We emphasize that neither Eq.~\eqref{eq:fT} nor its independence of the pressure model is claimed here as new. The triple-product condition was first obtained for magnetohydrostatic equilibria \cite{simakov_helander_2011}; necessary and sufficient conditions for QS were derived from the guiding-center Lagrangian without any equilibrium assumption in Ref.~\cite{rodriguez_necessary_2020}, and the triple-product form was recovered from them, again without assuming ideal equilibrium, in Ref.~\cite{rodriguez_magnetic}. An independent geometric characterization is given in Ref.~\cite{burby2020}. The point of rederiving the conditions here is that the symmetry in question is a symmetry of the full field--particle action of Sec.~\ref{sec:action}, the same functional whose Noether current gives the stress tensor. This is a point of some consequence, given that the central claim of this paper is that the scalar-pressure model itself is the problematic assumption in three dimensions.}
\end{addblock}

\begin{addblock}
\subsection{Cross-surface current, Boozer coordinates, and the parallel current}
\label{sec:parallel_current}

\add{The parallel force balance being an identity, the nontrivial content of the equilibrium problem resides in the perpendicular balance together with charge conservation. We now examine that content for an arbitrary kinetic equilibrium $f_{0s} = F_s(\Hgcs,\mu_s)$ with $\Phi = \Phi(\psi)$, for which, by Sec.~\ref{sec:parallel_identity}, $\ppar$, $\pperp$ and $\Dani$ are functions of $(\psi,B)$ satisfying Eq.~\eqref{eq:CGLidentity}.}

\add{Taking $\bB\times$ the force balance $\nabla\cdot\bPi = \bJ\times\bB/c$ and using Eq.~\eqref{eq:divPi} gives the perpendicular current}
\begin{equation}
  \bJ_\perp = \frac{c}{B^{2}}\,\bB\times\!\left(\nabla\pperp + \Dani\,\bkappa\right),
  \label{eq:Jperp}
\end{equation}
\add{the term along $\bhat$ in Eq.~\eqref{eq:divPi} having dropped out of the cross product. Charge conservation $\nabla\cdot\bJ = 0$ then yields the magnetic differential equation (MDE) for the parallel current,}
\begin{equation}
  \bB\cdot\nabla\!\left(\frac{\Jpar}{B}\right) = -\,\nabla\cdot\bJ_\perp \;\equiv\;S .
  \label{eq:MDE}
\end{equation}
\add{Equation~\eqref{eq:MDE} is the anisotropic generalization of the Pfirsch--Schl\"uter problem, and its solvability is the crux of the equilibrium existence question in three dimensions \cite{grad1967,newcomb1959,Helander2014-ev}. When $S$ is expanded in Fourier harmonics $e^{i(M'\theta - N'\phi)}$ in straight-field-line coordinates, the corresponding harmonic of $\Jpar/B$ acquires the denominator $\iota M' - N'$; in a generic 3D field the source contains all harmonics, so the solution is obstructed on a dense set of rational surfaces and suffers small divisors elsewhere \cite{grad1967,newcomb1959}.}

\paragraph{Current across the flux surfaces}
\add{Before turning to QS we note a property of Eq.~\eqref{eq:Jperp} that has no scalar-pressure counterpart. From $(\nabla\times\bB)\times\bB = (\bB\cdot\nabla)\bB - \nabla(B^2/2)$ and Amp\`ere's law, the curvature satisfies}
\begin{equation}
  \bkappa = \frac{\nabla_\perp B}{B} + \frac{4\pi}{c}\,\frac{\bJ\times\bB}{B^{2}} ,
  \label{eq:kappa_id}
\end{equation}
\add{so that $(\bB\times\bkappa)\cdot\nabla\psi = (\bB\times\nabla B)\cdot\nabla\psi/B + (4\pi/c)\,\bJ_\perp\cdot\nabla\psi$. Substituting in Eq.~\eqref{eq:Jperp}, writing $\nabla\pperp = (\partial\pperp/\partial\psi)\nabla\psi + (\partial\pperp/\partial B)\nabla B$, and using Eq.~\eqref{eq:CGLidentity} in the form $\partial\pperp/\partial B + \Dani/B = \partial(\ppar+\pperp)/\partial B$, we obtain}
\begin{equation}
  \bJ\cdot\nabla\psi \;=\; \frac{c}{B^{2}\,\sigma_f}\,
  \left.\frac{\partial(\ppar+\pperp)}{\partial B}\right|_{\psi}
  (\bB\times\nabla B)\cdot\nabla\psi ,
  \qquad
  \sigma_f \equiv 1 - \frac{4\pi\Dani}{B^{2}} ,
  \label{eq:Jdotgradpsi}
\end{equation}
\add{valid away from the firehose threshold $\sigma_f = 0$. The geodesic factor $(\bB\times\nabla B)\cdot\nabla\psi$ is generically nonzero in a toroidal field. Hence, in an anisotropic kinetic equilibrium, current crosses the flux surfaces locally unless $\ppar+\pperp$ is independent of $B$ on each surface. For scalar pressure the right-hand side vanishes identically.}

\paragraph{The averaged solvability condition is automatic under QS}
\add{The net current through a closed flux surface vanishes by Amp\`ere's law and Stokes' theorem, $\langle\bJ\cdot\nabla\psi\rangle_\psi = 0$, where $\langle\cdot\rangle_\psi$ is the flux-surface average. Since $\langle \bB\cdot\nabla g\rangle_\psi = 0$ for any single-valued $g$, and}
\begin{equation}
  \langle \nabla\cdot\bJ_\perp\rangle_\psi = \frac{1}{V'}\frac{d}{d\psi}\Bigl(V'\,\langle \bJ_\perp\cdot\nabla\psi\rangle_\psi\Bigr),
  \label{eq:divJperp_avg}
\end{equation}
\add{with $V'(\psi)$ the differential volume, this is equivalent, given regularity on the magnetic axis, to the solvability condition $\langle S\rangle_\psi = 0$ of Eq.~\eqref{eq:MDE}. In a quasisymmetric field this condition places no restriction on the anisotropy. From Eqs.~\eqref{eq:QS_cond} and \eqref{eq:QS_sympl} with $\Phi_\xi = \psi$ one has $\bB\times\nabla\psi = \bB(\bB\cdot\bxi) - B^2\bxi$, and therefore}
\begin{equation}
  (\bB\times\nabla B)\cdot\nabla\psi = -\,(\bB\cdot\bxi)\;\bB\cdot\nabla B .
  \label{eq:twoterm}
\end{equation}
\add{When $\bB\cdot\bxi$ is a flux function --- as it is in Boozer coordinates, where $\bB\cdot\bxi = (MG+NI)/(\iota M - N)$ for the normalization $\bB\times\bxi = \nabla\psi$ --- Eq.~\eqref{eq:twoterm} is the familiar two-term form of QS, and Eq.~\eqref{eq:Jdotgradpsi} becomes}
\begin{equation}
  \bJ\cdot\nabla\psi = -\,c\,(\bB\cdot\bxi)\;\bB\cdot\nabla W(\psi,B),
  \qquad
  W(\psi,B) \equiv \int^{B}\! \frac{1}{B'^{2}\sigma_f}\,\frac{\partial(\ppar+\pperp)}{\partial B'}\,dB' .
  \label{eq:Jdotgradpsi_QS}
\end{equation}
\add{Its flux-surface average vanishes identically for every $F_s(\Hgcs,\mu_s)$. The averaged condition is therefore not where QS constrains the anisotropy. In a generic 3D field, by contrast, where the ratio of $(\bB\times\nabla B)\cdot\nabla\psi$ to $\bB\cdot\nabla B$ is not a flux function, it is a genuine constraint.}

\paragraph{Boozer coordinates constrain the anisotropy}
\add{The binding condition is instead pointwise. The Boozer representation $\bB = G\nabla\phi + I\nabla\theta + \beta\nabla\psi$ with $G = G(\psi)$ and $I = I(\psi)$, on which the standard form~\eqref{eq:QS_def} of QS and the reduction below rest, requires $\bJ\cdot\nabla\psi = 0$ \cite{Helander2014-ev}. By Eq.~\eqref{eq:Jdotgradpsi}, and since $\bB\cdot\nabla B \not\equiv 0$ on a surface, this holds if and only if $\partial(\ppar+\pperp)/\partial B|_\psi = 0$, i.e., $\ppar + \pperp = 2p(\psi)$. Combining this with Eq.~\eqref{eq:CGLidentity}, which now reads $\partial\ppar/\partial B = 2(\ppar - p)/B$, gives}
\begin{equation}
  \ppar = p(\psi) + K(\psi)\,B^{2},
  \qquad
  \pperp = p(\psi) - K(\psi)\,B^{2},
  \qquad
  \Dani = 2K(\psi)\,B^{2},
  \qquad
  \sigma_f = 1 - 8\pi K(\psi).
  \label{eq:pK}
\end{equation}
\add{Kinetic equilibria compatible with Boozer coordinates therefore carry exactly one free flux function, $K(\psi)$, beyond the scalar pressure $p(\psi)$, and scalar pressure is recovered for $K = 0$. The mean of the two pressures being a flux function is also the structure of the anisotropic equilibria obtained by imposing QS on the Kruskal--Kulsrud functional through a Lagrange multiplier \cite{rodriguez_energy_principle}; here it is a consequence of the kinetic moments, with no multiplier. Which functions $p(\psi)$ and $K(\psi)$ can be realized by a non-negative $F_s(\Hgcs,\mu_s)$ is a well-posed inverse problem of the type familiar from mirror equilibrium theory \cite{taylor1963}. We emphasize that the weak QS conditions~\eqref{eq:QS_cond} and \eqref{eq:QS_sympl} do not themselves require Boozer coordinates; kinetic equilibria outside the class~\eqref{eq:pK} are not excluded by QS, but they carry a local cross-surface current~\eqref{eq:Jdotgradpsi_QS} and cannot be described in the standard Boozer form.}

\paragraph{The parallel current in closed form}
\add{For the class~\eqref{eq:pK}, $\nabla\pperp + \Dani\bkappa = (p' - K'B^2)\nabla\psi - 2KB\nabla B + 2KB^2\bkappa$, where primes denote $d/d\psi$. Using Eq.~\eqref{eq:kappa_id}, the terms in $\nabla_\perp B$ cancel and the term in $\bJ\times\bB$ reproduces $\bJ_\perp$, so that Eq.~\eqref{eq:Jperp} becomes}
\begin{equation}
  \bJ_\perp = \frac{c\,\bigl(p' - K' B^{2}\bigr)}{1 - 8\pi K}\;\frac{\bB\times\nabla\psi}{B^{2}} .
  \label{eq:Jperp_K}
\end{equation}
\add{Since $\nabla\cdot(\bB\times\nabla\psi) = (4\pi/c)\bJ\cdot\nabla\psi = 0$ and the coefficients depend on $\psi$ only, the $K'$ term is divergence-free, and}
\begin{equation}
  S = -\nabla\cdot\bJ_\perp
  = \frac{2c\,p'}{1-8\pi K}\,\frac{(\bB\times\nabla\psi)\cdot\nabla B}{B^{3}} .
  \label{eq:S_K}
\end{equation}
\add{In a generic 3D field this source has the resonant structure described above. In a quasisymmetric field, Eq.~\eqref{eq:twoterm} turns it into a parallel derivative, $S = -c\,p'(\bB\cdot\bxi)\,(1-8\pi K)^{-1}\,\bB\cdot\nabla B^{-2}$, and the MDE~\eqref{eq:MDE} integrates immediately:}
\begin{equation}
  \frac{\Jpar}{B} = -\,\frac{c\,p'(\psi)}{1 - 8\pi K(\psi)}\;\frac{MG + NI}{\iota M - N}\;\frac{1}{B^{2}} \;+\; h(\psi),
  \label{eq:Jpar_closed}
\end{equation}
\add{where $h(\psi)$ is fixed by the net toroidal current on each surface. For $K = 0$ this is the scalar-pressure Pfirsch--Schl\"uter current of a quasisymmetric field \cite{Helander2014-ev}. The anisotropy enters through the factor $(1-8\pi K)^{-1}$, which diverges at the firehose threshold, and through the $K'$ term of Eq.~\eqref{eq:Jperp_K}, which modifies the perpendicular current and hence the field, but not $\Jpar$. The dense set of resonant obstructions of a generic 3D field is replaced by the \emph{single} resonance $\iota = N/M$ of the denominator. This is the precise sense in which QS restores integrability at the level of the equilibrium currents, and it substantiates the statement made in Sec.~\ref{sec:intro}.}

\add{We stress that this argument does not, by itself, re-derive or resolve the Garren--Boozer overdetermination, which is a statement about the joint counting of the QS conditions and force balance order by order in the distance from the magnetic axis \cite{garren_quasihelical_1991,garren_field_strength_1991}, and whose resolution by anisotropic pressure was established in Refs.~\cite{rodriguez_mhd_1,rodriguez_mhd_2}. What it does establish is the kinetic content of that resolution: within Boozer coordinates, the anisotropy admitted by a kinetic equilibrium is exactly $\Dani = 2K(\psi)B^2$, and anisotropy of any other form requires a local cross-surface current. Whether the single additional flux function $K(\psi)$ suffices to lift the overdetermination order by order is a question for the near-axis analysis.}

\paragraph{Rational surfaces}
\add{On a surface with $\iota = N/M$ the operator $\bB\cdot\nabla$ annihilates every single-helicity function, and Eq.~\eqref{eq:Jpar_closed} shows that the parallel current is singular there unless $p' = 0$. The quasisymmetric equilibrium problem therefore retains a residual resonance structure, but one of measure zero in $\psi$ and involving a single helicity rather than a dense set. A systematic treatment of this case, and its relation to the rigidity results of Ref.~\cite{constantin2021}, is left to future work.}
\end{addblock}

\section{Kinetic Stability of Quasisymmetric Equilibria}
\label{sec:stability}

\begin{addblock}
\add{The energy--Casimir method applied in this section is classical \cite{holm1985,morrison1987variational,morrison1998,morrison_pfirsch_1989}, and we make no claim of novelty for the general criterion. We include it because the central claim of this paper is that equilibrium and stability descend from one functional, and omitting the stability branch would leave that claim unsupported. What is specific to the present setting is that the QS momentum enters the free energy as an additional conserved functional, which determines which equilibria the method can reach; the dependence of $f_{0s}$ on $\Pxis$ beyond that, and the radial-gradient drive it carries, lies outside the criterion and is discussed in Sec.~\ref{sec:stability:trapped}.}
\end{addblock}

\subsection{Energy--Casimir stability principle}
\label{sec:stability:casimir}
We now derive a stability criterion for the anisotropic QS equilibrium within the same variational framework using the energy--Casimir method \cite{morrison1987variational}\add{; see Refs.~\cite{holm1985,morrison_pfirsch_1989,morrison_pfirsch_1990} for the general theory}. For an equilibrium distribution $f_{0s}$ satisfying $\{f_{0s}, \Hgcs\}_{\mathrm{gc}} = 0$, stability is determined by the sign of the second variation of the free energy:
\begin{equation}
  \delta^2\calF > 0
  \label{eq:stability}
\end{equation}
for all dynamically accessible perturbations $\delta f_s = -\{f_{0s}, g\}_{\mathrm{gc}}$. Following the derivation in Sec. \ref{sec:action}, the free energy is:
\begin{equation}
\label{eq:free_energy_1}
     \calF = \calH \addblk{+ \lambda\,\mathcal{P}_\xi} + \sum_{s} C_{s}.
\end{equation}
\add{Here $\mathcal{P}_\xi = \sum_s\int d^5Z\, f_s\Pxis$ is the total QS momentum, conserved because of the symmetry of Sec.~\ref{sec:QS}, and $\lambda$ is a constant. It must be included because $\Pxis$ is a dynamical invariant, not a Casimir of the guiding-center bracket: without it, the method reaches only equilibria that are independent of $\Pxis$ (cf.\ Sec.~\ref{sec:variational}).}
Writing the Casimir invariant $C_s$ in the form\add{, in which the dependence on $\mu_s$ is admissible because $\mu_s$ is a Casimir of the guiding-center bracket}:
\begin{equation}
    C_s = \int d^5Z\, G_s(f_s\addblk{,\mu_s}).
\end{equation}
Abbreviating the field contributions as $W_\text{field}$, $\calF$ becomes:
\begin{equation}
\label{eq:free_energy_2}
     \calF = \sum_{s} \left[\int d^5Z\,G_s(f_s\addblk{,\mu_s}) +\int d^5Z\,\addblk{\mathcal{E}_s}\,f_s\right]+  W_\text{field},
     \qquad \addblk{\mathcal{E}_s \equiv \Hgcs + \lambda\Pxis,}
\end{equation}
The first and second variations of $\calF$ are as follows:
\begin{gather}
      \delta\calF =  \sum_{s} \int d^5Z\,\left(\frac{d G_s}{d f_s}+ \addblk{\mathcal{E}_s}\right)\delta f_s \addblk{+ \sum_s\int d^5Z\, f_s\,\delta_{\Phi,\bA}\mathcal{E}_s} + \delta W_\text{field} \label{eq:first_var}\\
      \delta^2\calF = \sum_{s} \int d^5Z\,\frac{d^2 G_s}{d f_s^2}(\delta f_s)^2 \addblk{+ \delta^2\calF_{\times}} + \delta^2 W_\text{field} \label{eq:second_var}
\end{gather}
\add{where $\delta_{\Phi,\bA}$ denotes the variation with respect to the potentials (including that of the measure), and $\delta^2\calF_{\times} = \sum_s\int d^5Z\,\bigl[2\,\delta f_s\,\delta_{\Phi,\bA}\mathcal{E}_s + f_s\,\delta^2_{\Phi,\bA}\mathcal{E}_s\bigr]$ collects the cross terms, arising for example from $\mu_s\,\delta B$.}
\begin{addblock}
\add{Because the phase-space variation $\delta f_s$ and the variations of the electromagnetic potentials are independent, the vanishing of $\delta\calF$ in Eq.~\eqref{eq:first_var} requires the two contributions to vanish separately. The field variation gives}
\begin{equation}
  \sum_s\int d^5Z\, f_s\,\delta_{\Phi,\bA}\mathcal{E}_s + \delta W_{\mathrm{field}} = 0,
  \label{eq:dWfield}
\end{equation}
\add{which is the statement that the equilibrium fields satisfy Gauss's law and Amp\`ere's law with the equilibrium charge and current densities, magnetization current included, as already obtained in Sec.~\ref{sec:action} by direct variation of the action. (The particle term is essential: $\delta W_{\mathrm{field}} = 0$ alone would give the vacuum equations.) The phase-space variation gives Eq.~\eqref{eq:casimir_eq} below. The field contributions do not disappear from the stability problem. $\delta^2 W_{\mathrm{field}}$ is positive definite for perturbations that leave the boundary fixed, but the cross terms $\delta^2\calF_\times$ are indefinite, so that positivity of the kinetic and field-energy terms separately does not by itself establish positivity of $\delta^2\calF$. The cross terms are absent for electrostatic perturbations of an unmagnetized Vlasov--Poisson plasma, for which the method was originally formulated \cite{morrison1987variational}; in the magnetized case they must be bounded, as in the free-energy analyses of Refs.~\cite{morrison_pfirsch_1989,morrison_pfirsch_1990}.}
\end{addblock}
At the equilibrium $f_{0s}$, $\delta\calF=0$, which gives
\begin{equation}
\label{eq:casimir_eq}
   \left.\frac{d G_s}{df_s}\right|_{f_{0s}}=-\left.\Hgcs\right|_{f_{0s}} \addblk{-\lambda\left.\Pxis\right|_{f_{0s}}}.
\end{equation}
\del{Assuming $f_{0s}$ has the form $F_s(\Hgcs,\mu_s,\Pxis)$ in \eqref{eq:equil_QS}, $F_s$ can be inverted to express $\Hgcs$ as a function of $f_s$ at fixed $\mu_s,\Pxis$.} \add{Equation~\eqref{eq:casimir_eq} says that the equilibria reached by the method are those of the form $f_{0s} = F_s(\mathcal{E}_s,\mu_s)$, i.e., the subclass of Eq.~\eqref{eq:equil_QS} in which $\Hgcs$ and $\Pxis$ enter only through the combination $\mathcal{E}_s = \Hgcs + \lambda\Pxis$. For such equilibria $F_s$ can be inverted to express $\mathcal{E}_s$ as a function of $f_s$ at fixed $\mu_s$.} This allows us to differentiate both sides of \eqref{eq:casimir_eq} to yield:
\begin{equation}
   \left.\frac{\partial}{\partial f_s}\left(\left.\frac{\partial G_s}{\partial f_s}\right|_{F_s}\right)\right|_{F_s}=-\left.\frac{\partial\addblk{\mathcal{E}_s}}{\partial f_s}\right|_{F_s}=-\frac{1}{F_s'},
\end{equation}
where \del{$F_s'=\partial{F_s}/\partial\Hgcs$}\add{$F_s' = \partial F_s/\partial\mathcal{E}_s$ at fixed $\mu_s$}. Consequently, the second variation of $\calF$ is:
\begin{equation}
  \delta^2\calF
  = -\sum_s\int d^5Z\,\frac{1}{F_s'}\,(\delta f_s)^2
  \addblk{+ \delta^2\calF_{\times}}
  + \delta^2W_{\mathrm{field}}.
  \label{eq:2nd_var}
\end{equation}
\del{A sufficient condition for stability is} \add{Positivity of the kinetic term requires}
\begin{equation}
  -F_s'\addblk{(\mathcal{E}_s,\mu_s)} > 0,
  \label{eq:mono_cond}
\end{equation}
which generalizes the monotonicity condition for Vlasov equilibria\add{; it is a sufficient condition for stability when, in addition, the cross terms $\delta^2\calF_\times$ are dominated by the positive terms, as they are trivially in the electrostatic, unmagnetized case}. \del{More generally, for quasisymmetric equilibria, a sufficient condition for stability is} \add{For quasisymmetric equilibria the condition reads}
\begin{equation}
  \left.\frac{\partial f_{0s}}{\partial \Hgcs}\right|_{\mu_s,\Pxis} \addblk{= \frac{\partial F_s}{\partial\mathcal{E}_s}} < 0,
  \label{eq:kin_Newcomb}
\end{equation}
which plays the role of a kinetic Newcomb criterion \del{and ensures the positive definiteness of} \add{for the kinetic part of} $\delta^2\calF$. \add{Two limitations should be stated plainly. First, the criterion applies only to the subclass $F_s(\mathcal{E}_s,\mu_s)$; a general dependence on $\Pxis$ is not a stationary point of $\calF$ for any choice of $G_s$ and $\lambda$, and its stability is not addressed by this method. Second, monotonicity in the energy alone is not a sufficient condition for stability of a distribution that also depends on $\Pxis$, i.e., on $\psi$: a distribution that is Maxwellian in energy with a radial density gradient satisfies Eq.~\eqref{eq:kin_Newcomb} and yet drives drift-wave and trapped-particle instabilities. Gardner-type stability results require the distribution to be a decreasing function of the energy alone, or of the energy in a uniformly rotating frame, which is the role played here by $\mathcal{E}_s$.} \add{In the appropriate limits this criterion reduces to the Kruskal--Oberman energy principle \cite{kruskal_oberman_1958} and, for isotropic distributions, to the classical magnetohydrodynamic energy principle \cite{bernstein1958}.}

\subsection{Trapped-particle effects and resonance}
\label{sec:stability:trapped}

In toroidal systems, trapped particles experience bounce motion and may resonate with perturbations, producing kinetic instabilities when $\omega = m_r\omega_{b} + n_r\omega_{d}$, where $\omega_{b}$ and $\omega_{d}$ are bounce and drift frequencies, and $m_r$ and $n_r$ are resonance mode numbers \add{\cite{kadomtsev_pogutse_1971}}. \del{Such instabilities require violations of the monotonicity condition~\eqref{eq:kin_Newcomb} with respect to the relevant invariants.} \add{Such instabilities are not excluded by Eq.~\eqref{eq:kin_Newcomb}: their free energy resides in the dependence of the distribution on the invariants other than the energy, in particular in the radial gradients, which in a quasisymmetric field appear as the dependence on $\Pxis$.} Trapped-electron modes correspond to regions where $\partial f_{0s}/\partial \Pxis$ changes sign or becomes sufficiently large.

QS alone does not eliminate all kinetic instabilities. However, the additional invariant $\Pxis$ modifies the orbit structure and resonance conditions, potentially reducing the resonant drive relative to generic 3D fields. \add{This is the kinetic counterpart of the structural point made throughout this paper: in a generic 3D field an exact equilibrium distribution can depend only on $(\Hgcs,\mu_s)$, whereas quasisymmetry supplies $\Pxis$ as a global invariant, so that the radial-gradient drive is carried by a well-defined derivative $\partial f_{0s}/\partial\Pxis$ of an exact equilibrium. That derivative is not constrained by the energy--Casimir criterion above, and a stability analysis of the drive requires the resonant, orbit-averaged treatment indicated here. A related resilience of optimized stellarators against trapped-particle instabilities has been established for quasi-isodynamic configurations \cite{proll2012,helander_proll_plunk_2013}, and it would be of interest to know whether the mechanism identified there admits an interpretation within the present variational structure.} In appropriate limits, the kinetic stability conditions derived here reduce to the CGL and MHD energy principles, providing a unified treatment of kinetic and fluid stability within a single variational framework.

\section{Implications for Equilibrium Construction and Magnetic Configuration Design}
\label{sec:design}

\begin{addblock}
\subsection{A variational measure of quasisymmetry breaking}
\label{sec:qs_metric}

\add{Because QS has been identified as a continuous symmetry of the guiding-center action, its failure admits a canonical measure: the rate at which the associated Noether momentum fails to be conserved. Some care is needed in choosing the vector fields $\bxi$ over which this rate is evaluated. If $\bxi$ were required only to be tangent to the flux surfaces, the choice $\bxi\propto\nabla\psi\times\nabla B$ would make $\bxi\cdot\nabla B$ vanish identically in any field, and any residual built on it would vanish for every configuration. The momentum $\Pxi$ of Eq.~\eqref{eq:Pxi} generates the flow along $\bxi$ only when the symplectic conditions~\eqref{eq:QS_sympl} hold, and these are what give the residual content. We therefore define, on each flux surface, the admissible class}
\begin{equation}
  \mathcal{X}_\psi = \bigl\{\,\bxi \;:\; \nabla\cdot\bxi = 0,\ \ \bB\times\bxi = \nabla\psi \,\bigr\}.
  \label{eq:Xadm}
\end{equation}
\add{The second condition fixes the perpendicular part, $\bxi_\perp = \nabla\psi\times\bB/B^2$, and the first then determines the parallel part $\alpha\bB$ through the magnetic differential equation $\bB\cdot\nabla\alpha = -\nabla\cdot(\nabla\psi\times\bB/B^2)$, up to a flux function. In Boozer coordinates $\mathcal{X}_\psi$ contains, for each helicity, the field $\bxi = (N\partial_\theta + M\partial_\phi)/(\iota M - N)$. For $\bxi\in\mathcal{X}_\psi$, at leading order in $\epsilon$,}
\begin{equation}
  \dot{\Pxi} = \{\Pxi, \Hgc\}_{\mathrm{gc}} = -\,\bxi\cdot\nabla\Hgc
    = -\left(\mu\,\bxi\cdot\nabla B + q\,\bxi\cdot\nabla\Phi\right),
  \label{eq:Pxidot}
\end{equation}
\add{and the second term vanishes for $\Phi = \Phi(\psi)$. This suggests the residual functional}
\begin{equation}
  \calQ(\psi) \;=\;
  \min_{\bxi\in\mathcal{X}_\psi}\;
  \frac{\displaystyle \sum_s \left\langle \int d^2V\, f_{0s}\,
   \left(\bxi\cdot\nabla \Hgcs\right)^{2} \right\rangle_\psi}
  {\displaystyle \langle|\bxi|^2\rangle_\psi\, L^{-2}\sum_s \left\langle \int d^2V\, f_{0s}\, \Hgcs^{2}
   \right\rangle_\psi } ,
  \label{eq:Qmetric}
\end{equation}
\add{where $L$ is a macroscopic length and the factor $\langle|\bxi|^2\rangle_\psi$ renders $\calQ$ independent of the normalization of $\bxi$. By construction, $\calQ(\psi) = 0$ if and only if the weak QS conditions~\eqref{eq:QS_cond} and \eqref{eq:QS_sympl} hold on the surface $\psi$.}

\add{Two features distinguish Eq.~\eqref{eq:Qmetric} from the residuals in common use. First, it does not require Boozer coordinates, which, as shown in Sec.~\ref{sec:parallel_current}, need not exist for an anisotropic equilibrium, and the helicity is not specified in advance but is selected by the minimization over $\mathcal{X}_\psi$; in practice the minimization reduces to the choice of the free flux function in the parallel part of $\bxi$ and, in a near-QS field, of the helicity. Second, it is weighted by the equilibrium distribution function, so that symmetry breaking is penalized in proportion to the phase-space density of particles that actually experience it, whereas a purely geometric residual constructed from the Boozer spectrum of $B$ weights all regions of the surface and all pitch angles equally. For a Maxwellian $f_{0s}$ we expect Eq.~\eqref{eq:Qmetric} to reduce to a temperature-weighted measure of the variation of $B$ along the candidate symmetry direction, closely related to the conventional two-term residual; we have not verified this in detail.}

\add{The quantity that governs the radial excursion of a trapped particle is, however, not the pointwise residual but its bounce average, since the secular change of $\Pxi$ over a bounce period is $\Delta \Pxi \simeq -\oint dt\,(\mu\,\bxi\cdot\nabla B)$. Replacing the integrand of Eq.~\eqref{eq:Qmetric} by the square of the bounce-averaged quantity gives}
\begin{equation}
  \calQ_b(\psi) \;=\;
  \min_{\bxi\in\mathcal{X}_\psi}\;
  \frac{\sum_s \bigl\langle \int_{\mathrm{trapped}} d^2V\, f_{0s}\,
   \bigl(\overline{\bxi\cdot\nabla \Hgcs}\bigr)^{2} \bigr\rangle_\psi}
  {\displaystyle \langle|\bxi|^2\rangle_\psi\, L^{-2}\sum_s \left\langle \int d^2V\, f_{0s}\, \Hgcs^{2}
   \right\rangle_\psi },
  \label{eq:Qb}
\end{equation}
\add{with the overbar denoting the bounce average. $\calQ_b$ vanishes for quasisymmetric fields. Omnigenity is defined by the vanishing of the bounce-averaged radial drift \cite{cary_shasharina_1997,landreman_catto_2012}, and it is natural to ask whether $\calQ_b$ also vanishes for omnigenous fields that are not quasisymmetric; we have not established this, and it is not implied by the construction above, since the admissible class~\eqref{eq:Xadm} is tied to the symmetry rather than to the bounce-averaged drift. In this respect $\calQ_b$ may be compared with the $\epsilon_{\mathrm{eff}}$ and $\Gamma_c$ metrics \cite{nemov1999,nemov2008}, which are constructed from transport estimates; $\calQ_b$ is derived from the symmetry structure of the action and does not require solving a drift-kinetic equation.}
\end{addblock}

\subsection{Consequences for equilibrium construction}
\label{sec:design:construction}

The unified variational framework has direct implications for equilibrium construction. For stationary states, the kinetic equation~\eqref{eq:gc_Vlasov} requires the equilibrium distribution to depend only on invariants of motion [Eq.~\eqref{eq:equil_QS}], and macroscopic equilibrium must satisfy the anisotropic CGL force balance [Eq.~\eqref{eq:CGL}] with the pressure tensor~\eqref{eq:gyrotropic_P}. The QS conditions must also hold to ensure the existence of $\Pxi$ and the integrability of orbits. Thus, equilibrium construction must simultaneously satisfy guiding-center invariance, anisotropic force balance, and the QS solvability constraint. This is a coupled kinetic-magnetic problem fundamentally different from scalar-pressure MHD.

In the unified formulation, anisotropy arises from the conservation of $\mu$; parallel and perpendicular pressures [Eq.~\eqref{eq:pressures}] are generically distinct\add{, as Eq.~\eqref{eq:CGLidentity} makes precise}. Scalar-pressure equilibria represent a special restricted subclass in which the $\mu$-dependence of $F$ is tuned to eliminate anisotropy; \del{such tuning is structurally unstable and does not represent the generic kinetic state.} \add{in the collisionless theory this is a restriction rather than the generic case, although collisions drive the distribution toward it (Sec.~\ref{sec:collisionality}).} \del{This also explains why} \add{This is consistent with the finding \cite{rodriguez_mhd_1,rodriguez_mhd_2} that} scalar-pressure models are blocked by the Garren--Boozer conundrum at higher perturbative orders: in 3D configurations, QS requires the extra degrees of freedom provided by the anisotropic pressure difference $(\pperp - \ppar)$ to satisfy global equilibrium equations. \add{Section~\ref{sec:parallel_current} identifies what those degrees of freedom are for a kinetic equilibrium described in Boozer coordinates: a single flux function $K(\psi)$, with $\Dani = 2K(\psi)B^2$.} Equilibrium solvers that assume scalar-pressure may therefore fail to capture admissible solutions of the full kinetic variational problem.

\begin{addblock}
\add{This last observation has a practical corollary for optimization practice. Stellarator optimization is at present carried out almost entirely against scalar-pressure equilibria \cite{qs_opt_example_Paul2021-cy,Landreman2022-lm}. If, as argued above, the admissible solution branch of the full kinetic problem carries nonzero anisotropy, then a scalar-pressure solver is restricted to a subclass of that branch, and QS targets evaluated on such equilibria are being evaluated on a constrained solution manifold. Quantifying the size of this effect --- by comparing optimization against anisotropic and scalar-pressure equilibria for a fixed target --- is a concrete numerical experiment that the present framework makes well posed, and one that existing anisotropic equilibrium codes \cite{iacono1990,cooper2009} are in principle equipped to perform.}
\end{addblock}

\begin{addblock}
\subsection{Finite collisionality}
\label{sec:collisionality}

\add{The variational structure developed in this paper is exactly collisionless: collisions destroy both the Noether structure of the action and the adiabatic invariance of $\mu$ on which the guiding-center reduction rests. It is therefore worth stating the regime of validity explicitly, and asking what becomes of the anisotropic resolution as collisionality increases.}

\add{The framework requires, in addition to the drift-kinetic ordering \eqref{eq:ordering}, the collisionality ordering \eqref{eq:coll_ordering}, $\nu_{ss'} \ll \omega_b,\ \omega_t$ --- the long-mean-free-path regime characteristic of reactor-grade stellarator cores. (The symbols have their usual notations.) Pitch-angle scattering isotropizes the distribution at a rate of order $\nu$, so that a steady-state anisotropy reflects a balance between drive --- magnetic pumping associated with the variation of $B$ along field lines, the trapped-particle population, anisotropic heating, and parallel flows --- and collisional relaxation, giving schematically $\Dani/p_\perp \sim \gamma_{\mathrm{drive}}/\nu$. In the opposite limit $\nu_* \gg 1$ the distribution is driven toward a function of $\Hgcs$ alone and, by the identity \eqref{eq:CGLidentity}, $\Dani \to 0$: the description degenerates continuously to scalar-pressure MHD, as it must. An important open question follows: how large an anisotropy, measured in the class~\eqref{eq:pK} by $4\pi\Dani/B^2 = 8\pi K$, is needed to make a given quasisymmetric configuration accessible, and whether that level is consistent with the drive available at reactor-relevant collisionality.}

\add{We note finally that the transport benefit of QS is not itself a collisionless phenomenon: a quasisymmetric field is isomorphic to an axisymmetric one for the purposes of neoclassical transport \cite{boozer1983,Helander2014-ev}. The neoclassical theory of stellarators at finite collisionality is developed in Refs.~\cite{Helander2014-ev,helander_simakov_2008}. A systematic treatment requires adding a collision operator to the action framework, for which the metriplectic and double-bracket formulations \cite{morrison1986,coquinot_morrison_2020} provide a natural setting, and which we leave to future work.}
\end{addblock}

\section{Discussion and Conclusions}
\label{sec:conclusions}

This work \add{addresses}\del{resolves} a persistent conceptual gap in stellarator theory by demonstrating that QS and pressure anisotropy are not independent features but dual manifestations of a single kinetic variational principle. While classical approaches often treat QS as an external geometric constraint and anisotropy as an \textit{ad hoc} fluid closure, we derive both from a unified guiding-center Vlasov--Maxwell action. Macroscopic force balance $\bJ \times \bB /c= \nabla \cdot \bPi$ with the gyrotropic stress tensor $\bPi$ is a direct consequence of Noether's theorem applied to this action. The geometric constraint~\eqref{eq:QS_cond}\add{, together with the symplectic conditions~\eqref{eq:QS_sympl},} emerges as the solvability condition for the existence of the Noether invariant $\Pxi$, unifying geometric symmetry, kinetic confinement, and constrained variational formulations as different aspects of the same underlying Hamiltonian structure.

Our principal results are:
\begin{enumerate}
  \item The guiding-center Vlasov--Maxwell system admits a unified Eulerian
        variational formulation in which admissible variations are restricted to
        dynamically accessible phase-space transformations\add{, following Refs.~\cite{Littlejohn1983-wh,Brizard2016-ts}}.

  \item Macroscopic force balance follows from the spatial translation symmetry of
        the action and yields the anisotropic CGL equilibrium
        equation~\eqref{eq:CGL} directly from kinetic moments, without
        phenomenological closure.

  \item \add{Within the framework we recover, with an exact treatment of the
        guiding-center Jacobian, the classical result \cite{taylor1963,hall_mcnamara_1975}
        that the parallel projection of that balance is an identity of any
        equilibrium distribution built on the invariants of the reduced motion:
        $\partial \ppar/\partial B|_\psi = (\ppar-\pperp)/B$, Eq.~\eqref{eq:CGLidentity}.
        Pressure anisotropy is thus a kinematic consequence of $\mu$-conservation
        together with the variation of $B$ along field lines, and isotropy is
        generically a restriction on the collisionless distribution function.}

  \item QS is identified as a continuous symmetry of the guiding-center
        \del{Hamiltonian,} \add{Lagrangian, both its Hamiltonian and its symplectic part,} whose associated Noether invariant is the conserved
        generalized momentum $\Pxi$~\eqref{eq:Pxi}. \add{The solvability conditions
        for this symmetry, together with the requirement that the second adiabatic
        invariant be a flux function, reproduce the coordinate-free condition
        $\fT = 0$, without any assumption on the plasma equation of state, in
        agreement with the earlier equilibrium-free derivations of
        Refs.~\cite{rodriguez_necessary_2020,rodriguez_magnetic}; the new element is that
        the symmetry is one of the full field--particle action that also generates the stress tensor.}
        \del{The geometric constraint involving grad-psi and grad-B emerges as the solvability condition for this symmetry and for the second adiabatic invariant to be integrable. Constraints previously introduced in constrained variational formulations thus arise naturally from the unified action.}

  \item \add{In an arbitrary kinetic equilibrium the current crossing the flux
        surfaces is proportional to $\partial(\ppar+\pperp)/\partial B|_\psi$,
        Eq.~\eqref{eq:Jdotgradpsi}. Its flux-surface average vanishes identically in a
        quasisymmetric field, so the averaged solvability condition does not constrain
        the anisotropy there. The existence of Boozer coordinates, however, requires it to
        vanish pointwise, which restricts the pressures to
        $\ppar = p(\psi) + K(\psi)B^2$, $\pperp = p(\psi) - K(\psi)B^2$, Eq.~\eqref{eq:pK} ---
        consistent with the constrained Kruskal--Kulsrud formulation of
        Ref.~\cite{rodriguez_energy_principle}, but obtained here without a Lagrange
        multiplier. For this class the parallel current in a quasisymmetric field
        follows in closed form, Eq.~\eqref{eq:Jpar_closed}, with the single resonance
        $\iota = N/M$ in place of the dense set of resonant obstructions of a generic
        three-dimensional field.}

  \item \add{Because QS is a symmetry of an action, its breaking is measured by the
        non-conservation of $\Pxi$. Evaluated over the vector
        fields admitted by the symplectic conditions, this yields a
        distribution-weighted residual that does not require Boozer coordinates,
        Eq.~\eqref{eq:Qmetric}, and a bounce-averaged variant, Eq.~\eqref{eq:Qb},
        available as target functions for optimization without solving a
        drift-kinetic equation. Whether the bounce-averaged variant also vanishes
        for omnigenous fields is open.}

  \item Kinetic stability conditions follow from the energy--Casimir method and
        generalize Newcomb's energy principle to quasisymmetric guiding-center
        equilibria\add{. The general criterion is classical
        \cite{holm1985,morrison1987variational}; in the quasisymmetric case the
        conserved momentum must be added to the free energy, and the criterion
        then covers equilibria of the form $F_s(\Hgcs+\lambda\Pxis,\mu_s)$, subject to
        control of the field--particle cross terms. The radial-gradient drive
        carried by a more general dependence on $\Pxis$ lies outside it}.
\end{enumerate}

Several avenues for further work are open. The constructive implementation of anisotropic equilibrium solvers based on the unified action remains an important practical problem. Extensions to gyrokinetic theory would allow a systematic treatment of microinstabilities within the same variational structure, providing a foundation for microstability analysis using the Kruskal--Oberman energy principle \add{\cite{kruskal_oberman_1958}}. \del{Numerical optimization strategies for stellarator design may benefit from metrics derived directly from the symmetry properties of the guiding-center action --- specifically, the degree to which the action fails to possess the corresponding Noether symmetry, which provides a natural variational measure of QS breaking.} \add{The residual $\calQ$ introduced in Sec.~\ref{sec:qs_metric} should be evaluated on existing optimized configurations and compared with the geometric and transport-based metrics currently in use, both to validate it and to establish whether its distribution weighting materially changes the ranking of candidate configurations. The kinetic realizability of the class~\eqref{eq:pK} --- which profiles $p(\psi)$ and $K(\psi)$ arise from non-negative distributions --- and the size of $K$ needed to lift the Garren--Boozer overdetermination in near-axis expansions are natural next questions. The behavior of the theory at rational surfaces, where the parallel current~\eqref{eq:Jpar_closed} is singular unless $p'=0$, remains open, as does the systematic incorporation of collisions through a metriplectic extension of the action.}

\section*{Acknowledgements}
We acknowledge the use of ChatGPT and Claude (Anthropic) for writing assistance and literature search. This research is supported by the Department of Energy Award No. DE-SC0024548. 

\section*{Competing Interests}
We declare no competing interests.

\section*{Data Availability}
No new experimental or computational data were generated.

\bibliographystyle{unsrtnat}
\bibliography{references}

\end{document}